\documentclass[prl,twocolumn,superscriptaddress,showpacs]{revtex4-2}

\usepackage{amsmath,amssymb,amsfonts}
\usepackage{graphicx}
\usepackage{hyperref}
\usepackage{bm}
\usepackage{color}

\begin{document}

\title{Separation of even-even from even-odd isotopes
using ultrafast lasers}

\author{Jacob Levitt}
\email{jakelevitt@cortexfusion.systems}
\affiliation{Cortex Fusion Systems, Inc., New York, NY 10177, USA}

\date{\today}

\begin{abstract}
We propose a laser isotope separation mechanism in
which selectivity arises from nuclear spin rather
than isotope shifts, enabling the use of broadband
ultrafast lasers.  A Ramsey pulse sequence is applied
to paramagnetic molecular isotopologues possessing
two electronic states coupled by a dipole transition.
For even-even isotopologues (nuclear spin $I = 0$),
each electronic state is a single level and the
time-reversed sequence returns all population to the
ground state exactly.  For even-odd isotopologues
($I > 0$), the hyperfine interaction splits each
state into multiple levels with coupling amplitudes
set by Wigner $6j$ symbols; incommensurate phase
evolution during the dark interval prevents the echo
from closing, trapping a fraction $P_m$ of the
population in the excited manifold.  In the impulsive
limit ($\Omega \gg A_{\rm HF}$), $P_m$ depends only
on the angular momentum quantum numbers
$(J_g, J_m, I)$ and is independent of laser intensity
or bandwidth.  Density matrix simulations confirm
$P_m = 0$ for $I = 0$ and
$P_m \approx 0.23$--$0.47$ for $I > 0$ across
representative systems including
${}^{235}$U, ${}^{87}$Sr, and ${}^{57}$Fe.  Under
realistic collisional conditions, single-pass
enrichment exceeding 90\% from natural feed is
achievable without cascading.
\end{abstract}
\maketitle

Even-even isotopes
($Z \in 2\mathbb{Z}$, $N \in 2\mathbb{Z}$; nuclear spin
$I = 0$) have different applications from the corresponding even-odd isotopes
($Z \in 2\mathbb{Z}$, $N \in 2\mathbb{Z}+1$; $I \neq 0$)
of the same element. For example,
${}^{57}$Fe/${}^{54,56,58}$Fe in M\"{o}ssbauer spectroscopy;
${}^{87}$Sr/${}^{84,86,88}$Sr in optical lattice clocks;
${}^{91}$Zr/${}^{90,92,94,96}$Zr in nuclear fuel cladding;
${}^{113}$Cd/${}^{110,112,114,116}$Cd in reactor control rods;
${}^{129,131}$Xe/${}^{128,130,132,134,136}$Xe in anesthesia; and
${}^{235}$U/${}^{234,236,238}$U in nuclear fuel enrichment.

Traditional laser isotope separation relies on spectroscopically
resolving isotope shifts arising from differences in nuclear mass
and charge radius, which rarely exceed a few GHz even for
heavy elements. In the most studied case of uranium, AVLIS selectively
photoionizes ${}^{235}$U by tuning dye lasers to the
$\sim$10~GHz field shift between ${}^{235}$U and
${}^{238}$U atomic transitions~\cite{Paisner1988},
while the molecular schemes MLIS~\cite{Jensen1982},
SILEX~\cite{Snyder2016}, and
CRISLA~\cite{Eerkens1998} resolve the
$\sim$0.6~cm$^{-1}$ vibrational isotope shift of the
UF$_6$ $\nu_3$ mode. 

All four methods have been limited by two structural
tradeoffs that have prevented industrial deployment.
First, selectivity requires that the isotope-specific
excitation survive long enough to be exploited by
the separation step (photoionization, dissociation,
or chemical reaction). Because the selectivity resides in an isotope-specific
excited-state population, collisional quenching and
resonant energy transfer between isotopologues erase it
on a timescale that decreases with gas density, while
throughput demands high density. 
Second, the laser linewidth must remain below the
isotope shift ($\sim$0.5--10~GHz depending on the
scheme), while simultaneously delivering the average
power and repetition rate needed for industrial
throughput.  No commercial laser platform satisfies
these requirements simultaneously at the relevant
wavelengths, and the bespoke systems developed for
AVLIS and SILEX have not achieved the reliability
or cost targets needed for deployment.
\begin{figure*}[t]
  \centering
  \includegraphics[width=\textwidth]{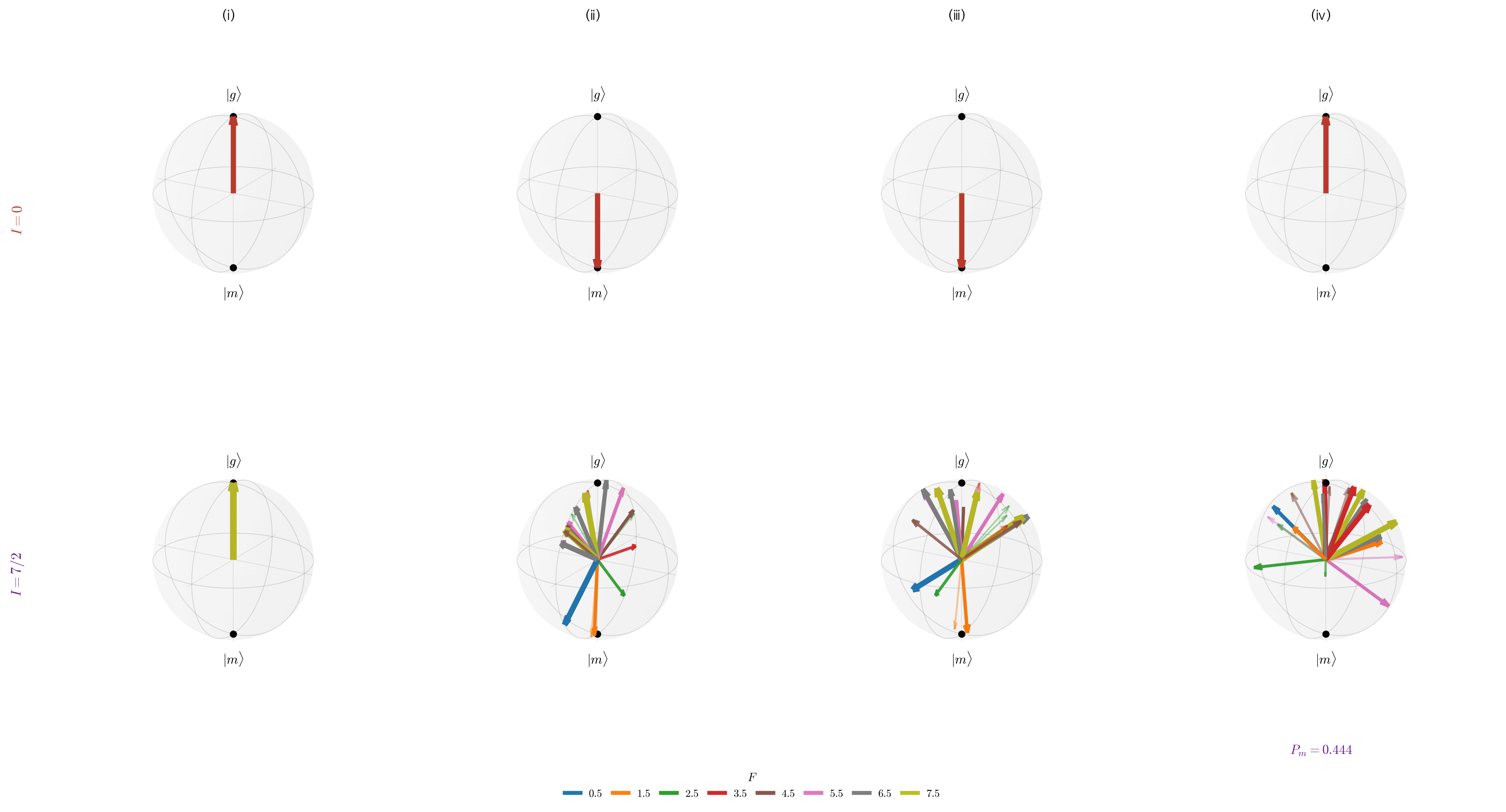}
  \caption{Bloch-sphere representation of the
  $\pi_0$--$\tau$--$\pi_\phi$ sequence for
  $J_g = 4$, $J_m = 5$.  Top row ($I = 0$), single
  channel, exact echo.  Bottom row ($I = 7/2$), all
  21 $(F, F')$ channels shown as unit Bloch vectors
  with line width encoding channel population.  Color
  indicates the starting ground-state $F$.
  Columns show (i)~initialization,
  (ii)~after first $\pi$ pulse,
  (iii)~after dark interval $\tau = 8$~ns,
  (iv)~after second $\pi$ pulse.}
  \label{fig:bloch}
\end{figure*}

In this paper, we show how, conversely, broadband
(i.e., ultrafast) lasers with bandwidths exceeding
1~THz can achieve isotope separation even when the
laser bandwidth exceeds the isotope shift by orders
of magnitude, by exploiting a selectivity mechanism
that is entirely independent of spectral resolution.  This opens isotope separation to the
workhorse ultrafast laser platforms already deployed
at industrial scale (e.g., mode-locked Ti:sapphire,
Yb:YAG, and fiber lasers), delivering high average
power at high repetition rates with the reliability,
cost, and maintenance characteristics of mature
commercial products.

\textit{Mechanism.}
Consider a paramagnetic molecule possessing two
electronic states $|g, J_g\rangle$ and $|m, J_m\rangle$
separated by energy $\omega_0$.  When the molecule
contains an even-even nucleus ($I = 0$), each
electronic state is a single sublevel and the two
states form an exact two-level system.  When the
molecule contains an even-odd nucleus ($I > 0$), the
coupling of nuclear spin $I$ to electronic angular
momentum $J$ splits each electronic state into
hyperfine levels $F$ with $|J - I| \leq F \leq J + I$.
The two-level system becomes an $(n_g + n_m)$-level
system threaded by $\Delta F = 0, \pm 1$ transition
pathways whose amplitudes are fixed by Wigner $6j$
symbols,
\begin{equation}
c_{FF'} = (-1)^{F' + J_g + I + 1}
\sqrt{(2F\!+\!1)(2F'\!+\!1)}\;
\begin{Bmatrix} J_g & J_m & 1 \\ F' & F & I
\end{Bmatrix}
\label{eq:6j}
\end{equation}

Now consider a broadband pulse coupling these states
with Rabi frequency $\Omega \gg A_{\rm HF}$.  The
pulse bandwidth far exceeds the hyperfine spread, so
all $F$ components are driven simultaneously.  However,
the $6j$ coefficients $c_{FF'}$ differ across channels,
so a pulse calibrated to area $\pi$ on the
root-mean-square coupling does not deliver area $\pi$
on any individual $(F, F')$ pair.  Each ground level
$F$ therefore distributes its population across
multiple metastable levels $F'$ in unequal proportions
set by the $6j$ algebra.

Applying the Ramsey sequence
$\pi_0$--$\tau$--$\pi_\phi$~\cite{Ramsey1950,Wynands2005},
where the second pulse is the time-reverse of the first
($U_2 = U_1^\dagger$), for $I = 0$, the echo closes
exactly and all population returns to the ground state,
$P_m = 0$, regardless of $\Omega$ or $\tau$.  For $I > 0$, the $6j$ asymmetry spreads population
across multiple metastable levels, and during the dark
interval $\tau$ each level evolves at its own hyperfine
frequency, scrambling the phase relationships that the
second pulse would need to recombine them.  Once $\tau \gg 1/A_{\rm HF}$, the cross-channel
coherences average to zero and the trapped population
in the excited manifold converges to a nonzero value
$P_m(J_g, J_m, I)>0$ that depends only on the angular
momentum quantum numbers.

\textit{Numerical demonstration.}
The Ramsey sequence was simulated by propagating the
thermal density matrix $\rho_0 = \sum_F w_F
|g,F\rangle\langle g,F|$ through the full
$(n_g + n_m)$-dimensional Hilbert space in the
rotating frame.  The driving Hamiltonian $H_{\rm drive}$
was constructed from the $6j$ coupling coefficients of
Eq.~\eqref{eq:6j} using exact Wigner symbol evaluation, and all propagators
$U_1 = \exp(-i H_{\rm drive}\, \pi/\Omega)$,
$U_\tau = \exp(-i H_{\rm free}\, \tau)$,
$U_2 = U_1^\dagger$ were computed by matrix
exponentiation in QuTiP~\cite{Johansson2012}.  The
metastable population was extracted as
$P_m = \mathrm{Tr}(\hat{P}_m\, \rho_f)$,
where $\hat{P}_m$ projects onto the metastable
manifold and $\rho_f = U_2 U_\tau U_1\, \rho_0\,
U_1^\dagger U_\tau^\dagger U_2^\dagger$.

Figure~\ref{fig:bloch} shows the resulting Bloch-sphere
trajectories for a representative system with
$J_g = 4$, $J_m = 5$ at $\Omega/2\pi = 50$~GHz and
$\tau = 8$~ns.  For $I = 0$ (top row), the single
Bloch vector executes a closed loop, returning exactly
to $|g\rangle$ after the second pulse.  For $I = 7/2$
(bottom row), each of the 21 coupled $(F, F')$
channels is plotted as an independent unit Bloch
vector with line width proportional to the population
in that two-level subspace.  After the first pulse,
the vectors fan across the metastable hemisphere with
unequal amplitudes reflecting the $6j$ coefficients;
during $\tau$, each vector precesses at its own
hyperfine frequency; and finally the second pulse cannot
simultaneously rephase all 21 channels, and the
ensemble retains $P_m = 0.444$ in the excited
manifold.

\begin{figure}
  \centering
  \includegraphics[width=\columnwidth]{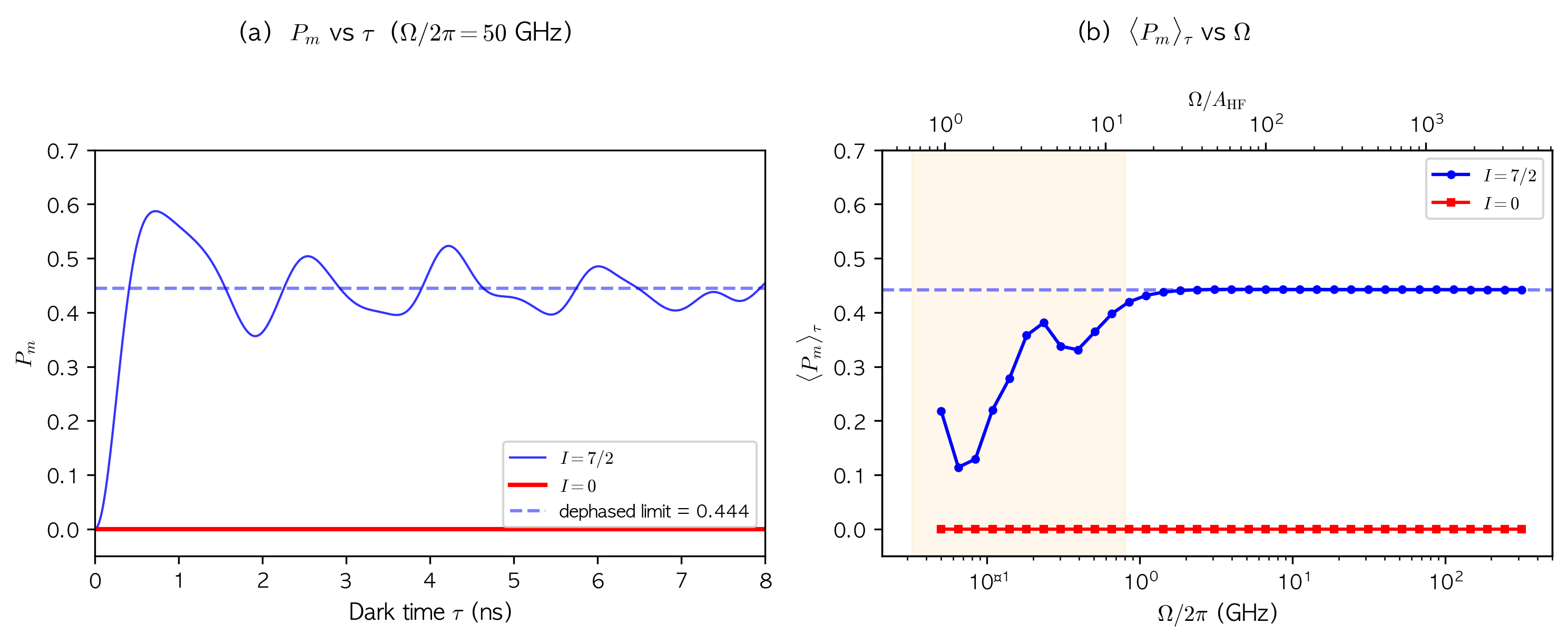}
  \caption{Invariance of $P_m$ for $J_g = 4$,
$J_m = 5$, $I = 7/2$.  (a)~$P_m$ vs dark
time $\tau$ at $\Omega/2\pi = 50$~GHz.  Red
($I = 0$) is identically zero.  Blue ($I = 7/2$)
rises from zero, oscillates on the timescale
$1/A_{\rm HF}$, and converges to 0.444 (dashed).
(b)~$\tau$-averaged $P_m$ vs Rabi frequency
$\Omega/2\pi$.  Shaded region indicates
$\Omega < 10\, A_{\rm HF}$ where the impulsive
limit breaks down.  Upper axis shows
$\Omega / A_{\rm HF}$.}
  \label{fig:independence}
\end{figure}

Figure~\ref{fig:independence} establishes the two
invariance properties of $P_m$.  In panel~(a),
$P_m$ is plotted against dark time $\tau$ at fixed
$\Omega/2\pi = 50$~GHz.  The $I = 0$ trace is zero at all $\tau$, while the $I = 7/2$
trace rises from zero as the hyperfine phases
develop, overshoots, and rings down to the
dephased limit $P_m = 0.444$ on the timescale
$\tau \sim 1/A_{\rm HF}$.  In panel~(b), the
$\tau$-averaged $P_m$ is plotted against
Rabi frequency $\Omega$.  Below
$\Omega / A_{\rm HF} \sim 10$, the impulsive
condition is violated and the trapping fraction
becomes intensity-dependent.  Above this threshold,
$P_m$ is flat across more than two decades
of $\Omega$, confirming that the result is
independent of laser power, pulse energy, and
focusing geometry.

Table~\ref{tab:pmeta} presents $P_m$ computed
from the same density matrix propagation across a range
of angular momentum quantum numbers spanning the
isotope pairs listed in the introduction.  In every
case, $P_m(I\!=\!0) = 0$
and $P_m(I\!>\!0)$ falls between 0.23 and
0.47.  Higher nuclear spin generally increases the
trapping fraction by opening more metastable channels,
though the dependence on $J_g$ and $J_m$ is
non-monotonic due to the oscillatory structure of the
$6j$ symbol.  The lowest trapping fractions occur for
$I = 1/2$, which couples each $F$ to only three
metastable levels and permits substantial recombination
even after dephasing.

\textit{Robustness to molecular structure.}
Three features of molecular dynamics that are absent
from the model merit discussion.

First, real molecules at finite temperature populate
many rotational levels $R$.  In the impulsive limit,
each rotational transition forms an independent
subspace with the same electronic quantum numbers
$(J_g, J_m, I)$, so the trapping fraction is identical
for every $R$ and the $I = 0$ echo closes channel by
channel.  The Rabi frequency varies across rotational
levels only through the fractional shift
$\delta\Omega/\Omega \sim 2B R / \Delta_{\rm int}$,
where $B$ is the rotational constant and
$\Delta_{\rm int}$ is the intermediate-state detuning.
For heavy paramagnetic molecules
($B \sim 0.01$~cm$^{-1}$,
$\Delta_{\rm int} \sim 10^4$~cm$^{-1}$),
this ratio is $< 10^{-5}$ even at $R = 100$, and the
values of $P_m$ in Table~\ref{tab:pmeta} would apply to
every populated rotational state without correction.

Second, the Ramsey dark interval
$\tau \sim 1/A_{\rm HF}$ (typically 1--100~ns) must be
shorter than the lifetime of the excited electronic
state.  For intra-configuration spin-orbit transitions
(e.g., $f$--$f$ or $d$--$d$), the radiative rate is
suppressed by the Laporte selection rule and the
small energy gap, giving natural lifetimes of
$10^{-3}$--$10^{3}$~s.  Nonradiative decay is
governed by the Franck-Condon overlap between the
two spin-orbit partners, which is extremely small
when the active electrons are core-like and the
equilibrium geometry does not change upon excitation.
The dark interval is therefore shorter than the
excited-state lifetime by many orders of magnitude
for any molecular platform in which the spin-orbit
states arise from a shielded electron configuration.

Third, vibronic decoherence, which limits electronic
coherence lifetimes in most molecular systems to
tens of
femtoseconds~\cite{Kaufman2023,Suchan2025},
is strongly suppressed for spin-orbit partner
states of the same electron configuration.
Because the two states share identical orbital
occupancy and differ only in the coupling of
angular momenta, their potential energy surfaces
are parallel along all nuclear
coordinates.  The Rossky-Bittner decoherence
rate~\cite{Rossky1995}, which scales as the
squared difference of gradients between surfaces,
is therefore expected to be negligible on the
nanosecond timescale of the dark interval
required for hyperfine dephasing.

\begin{table}[b]
  \caption{Trapping fraction $P_m(J_g, J_m, I)$
  computed by rotating-frame density matrix propagation
  in the impulsive, dephased limit.  The values of
  $J_g$ and $J_m$ are representative and depend on the
  molecular compound; $I$ is fixed by the nucleus.
  In all cases $P_m(I\!=\!0) = 0$ exactly.}
  \label{tab:pmeta}
  \begin{ruledtabular}
    \begin{tabular}{cccccc}
      $J_g$ & $J_m$ & $I$ & dim & $P_m$ &
      Example nuclei \\
      \hline
         1  &    2  & 1/2 &   4 & 0.368 &
      ${}^{57}$Fe, ${}^{113}$Cd, ${}^{129}$Xe \\
       5/2  &  7/2  & 1/2 &   4 & 0.270 &
      ${}^{57}$Fe, ${}^{113}$Cd, ${}^{129}$Xe \\
       3/2  &  5/2  & 3/2 &   8 & 0.420 &
      ${}^{131}$Xe \\
       7/2  &  9/2  & 3/2 &   8 & 0.403 &
      ${}^{131}$Xe \\
         2  &    3  & 5/2 &  11 & 0.470 &
      ${}^{91}$Zr \\
         3  &    4  & 7/2 &  15 & 0.467 &
      ${}^{235}$U \\
         4  &    5  & 7/2 &  16 & 0.444 &
      ${}^{235}$U \\
       9/2  & 11/2  & 7/2 &  16 & 0.428 &
      ${}^{235}$U \\
         5  &    6  & 9/2 &  20 & 0.454 &
      ${}^{87}$Sr \\
         6  &    7  & 9/2 &  20 & 0.431 &
      ${}^{87}$Sr \\
    \end{tabular}
  \end{ruledtabular}
\end{table}

\textit{Discussion.} Given a feed fraction $x_f$ of
the target isotope and a selectivity
$S = P_m(I\!>\!0) / P_m(I\!=\!0)$, the product
enrichment in a single pass is
\begin{equation}
x_p = \frac{S\, x_f}{(S - 1)\, x_f + 1}
\label{eq:enrich}
\end{equation}The residual $P_m(I\!=\!0)$ is set by three
mechanisms.  Rotational and vibrational detuning
errors scale as $(\sigma/\Omega)^2$ and are
suppressed in the deeply impulsive regime.  Beam
nonuniformity contributes as $(\delta I / I)^2$.
Collisional dephasing during the dark interval
sets a pressure-dependent floor that is independent
of $\Omega$ but decreases with decreasing gas
density.  The collisional contribution to $P_m(I\!=\!0)$ is
\begin{equation}
P_{\rm coll} = n\, \sigma_d\, v_{\rm rel}\, \tau
\label{eq:coll}
\end{equation}
where $n = P / k_B T$ is the molecular number
density, $\sigma_d$ is the dephasing cross section
(bounded above by the kinetic cross section),
$v_{\rm rel} = \sqrt{8 k_B T / \pi \mu}$ is the mean
relative speed, and $\tau \sim 1/A_{\rm HF}$ is the
dark interval.  Since $P_{\rm coll}$ scales linearly
with pressure and linearly with $\tau$, under realistic conditions (a few torr, room
temperature), $P_{\rm coll} \sim 10^{-4}$, giving
$S \sim P_m / P_{\rm coll} \sim 10^3$ and
single-pass enrichment exceeding 90\% from natural
${}^{235}$U feed (0.7\%) via
Eq.~\eqref{eq:enrich}. 

The Ramsey sequence deposits the even-odd
isotopologue into an electronic excited state
separated from the ground state by a spin-orbit
gap of order 0.1--1~eV.  Any subsequent separation
step inherits the broadband character of the
excitation, since the only spectral constraint is
that the separation laser remain below the
electronic gap.  With both steps broadband, the scheme maps onto commercial ultrafast laser
platforms.  Thin-disk oscillators now deliver 550~W
from a single modelocked
cavity~\cite{Seidel2024}, thin-disk multipass
amplifiers built for industrial deployment provide
1.9~kW at 10~mJ pulse energies~\cite{Dietz2020},
and coherently combined fiber systems have
surpassed 10~kW~\cite{Muller2020}, and optical parametric amplifiers providing
carrier-envelope-phase-stable coverage from the UV
to beyond 40~$\mu$m~\cite{Elu2021, Rothhardt2016}. Because the
entire excite-dephase-separate cycle completes in
$\sim$10--100~ns, collisional quenching can be
outrun even at static gas pressures of a few torr,
where the molecular number density is orders of
magnitude higher than in the effusive or supersonic
beams required by narrowband schemes.  The
combination of single-pass enrichment exceeding
90\%, gas-phase number densities of order
$10^{17}$~cm$^{-3}$, and repetition rates above
10~kHz places tonne-scale annual throughput within
reach of a single laser system.

We note that a physical implementation requires a
volatile paramagnetic compound with a spin-orbit
partner state accessible by dipole coupling
($\Delta J = 0, \pm 1$) and a state-selective
separation pathway.  These conditions are satisfied
by many $f$-block coordination compounds, where
spin-orbit splittings of 0.1--1~eV and volatile
molecular complexes are documented across the
lanthanide and actinide series.

\textit{Conclusion.}
We have shown that broadband ultrafast lasers can
separate even-even from even-odd isotopes by
exploiting the failure of a Ramsey echo in the
presence of molecular hyperfine structure.  The
trapped fraction $P_m$ is determined by the
angular momentum quantum numbers $(J_g, J_m, I)$
alone and is independent of laser intensity,
bandwidth, and interaction time in the
impulsive limit.  Because
selectivity derives from nuclear spin, both the
excitation and the separation steps are inherently
broadband, giving access to the high repetition
rates and high average powers of mature commercial
ultrafast laser platforms.  The mechanism applies
in principle to any even-$Z$ element possessing
both spin-zero and spin-nonzero stable isotopes.
By replacing spectral resolution with angular
momentum algebra as the basis of selectivity,
ultrafast laser isotope separation becomes a
problem not of precision spectroscopy but of
molecular quantum control.
\newline
\newline
\begin{acknowledgments}
The subject matter disclosed herein is covered by
U.S.\ Patent Application Nos.\
63/974,702 and 63/996,961 and related filings,
assigned to Cortex Fusion Systems, Inc.  All
commercial rights to the methods and systems
described in this work are exclusively held by
Cortex Fusion Systems, Inc.
\end{acknowledgments}

\end{document}